# Estimating the prevalence of malicious extraterrestrial civilizations

Alberto Caballero

alberto.caballero@uvigo.es

ABSTRACT   This paper attempts to provide an estimation of the prevalence of hostile extraterrestrial civilizations through an extrapolation of the probability that we, as the human civilization, would attack or invade an inhabited exoplanet once we become a Type-1 civilization in the Kardashev Scale capable of nearby interstellar travel. The estimation is based on the world's history of invasions in the last century, the military capabilities of the countries involved, and the global growth rate of energy consumption. Upper limits of standard deviations are used in order to obtain the estimated probability of extraterrestrial invasion by a civilization whose planet we send a message to. Results show that such probability is two orders of magnitude lower than the impact probability of a planet-killer asteroid. These findings could serve as a starting point for an international debate about sending the first serious interstellar radio messages to nearby potentially habitable planets.

Key-words: search for extraterrestrial intelligence, alien life, habitable exoplanets, kardashev scale, drake equation, fermi paradox.

Introduction

Active SETI (Active Search for Extra-Terrestrial Intelligence), sometimes referred as METI (Messaging to Extra-Terrestrial Intelligence), is a form of SETI that has been barely practised since the first interstellar radio message, the Arecibo Message, was sent in 1974 to the globular star cluster M13. Since then, there has been some attempts to contact potentially habitable exoplanets, such as Gliese 581 c, Tau Ceti f, and GJ 273 b. As of today, only GJ 273 b, located 19 light years away, remains considered to be potentially habitable, with an estimated earth similarity of 85%. The message, which was sent under a project called 'Sónar Calling Gliese 273 b', will arrive in 2029, and any response to Earth would be received as of 2041. The problem is that the message only contained music and a scientific tutorial to decode it. This is, it was a symbolic interstellar message, and therefore unlikely to be received and decrypted by any civilization provided that it exists.

The cause why there has been no serious attempts to send a radio message to a potentially habitable exoplanet is because there has been no international debate to discuss the possibility of doing it, and there has been no discussion because there is a prevailing fear in the scientific community that the message could be picked up by a malicious extraterrestrial civilization. Such civilization could be more advanced than ours, and perhaps interested in exploiting the resources of our planet by force.

For this reason, it becomes urgent to find an approximation to determine how likely it would be that an intelligent civilization living in the exoplanet we message has malicious intentions and poses a threat to humanity. Zero risk does not exist, and will never exist. However, if such risk is similar than that of other natural catastrophic events that could affect the entire Earth, then the advantages that would derive from finding and communicating with an intelligent civilization could greatly exceed its drawbacks. Ironically, this would be the case if the extraterrestrial civilization is far more advanced than ours, which could allow us to improve our technology and provide humanity with significant benefits.

## Methodology

The estimation of the probability of an intelligent civilization being malicious is based on the probability of humanity posing that same threat to them once it becomes a Type-1 civilization capable of nearby interstellar travel. To do so, a frequency distribution of the countries that have invaded others between 1915 and 2022 is conducted. A total of 51 countries are found to have invaded another country at some point during the referred time frame. The invaders in World War I and II are limited to those that initially started the invasions and triggered the rest of them. Those invasions that formed part of a large occupation plan including several countries are counted as a single invasion. Moreover, civil wars are not taken into account since they do not involve the invasion of any foreign territory.

The probability of invasion of each country is weighted with its percentage of global military expenditure, which can be considered as a variable of capability. With a weighted average mean [1], the individual probabilities of invasion are then added and divided by the total number of countries in order to estimate the current human probability of invasion of an extraterrestrial civilization.

$$W = \frac{\sum_{i=1}^{n} w_i X_i}{\sum_{i=1}^{n} w_i} \quad [1]$$

$$\sigma = \sqrt{\frac{\sum (x_i - \mu)^2}{N}} \quad [2]$$

$$f(x) = a(1+r)^x \quad [3]$$

With the upper limit obtained from the standard deviation [2] of the human probability of invasion and the average rates of global invasion as well as energy consumption, an exponential growth calculation [3] allows us to estimate the probability of invasion of a Type-1 civilization. The minimum estimated number of potentially habitable planets in our galaxy and the upper-limit of the standard deviation of the estimated number of intelligent civilizations in our galaxy allow us to obtain the probability that the exoplanet we message has complex life. This probability is then multiplied by the probability that any civilization has malicious intentions, which yields the total probability that the exoplanet we specifically send an interstellar message to is inhabited by a malicious civilization. Based on that same upper-limit , we can obtain the population of those that would be malicious and, therefore, likely to invade or attack the Earth.

Results

After calculating the weighted probabilities of invasion of each country [see figure 1], and considering that Earth has a total of 195 official countries, the estimated probability of invasion by a Type-0 civilization as humanity not capable of interstellar travel is 0.026%. However, given the threat that a malicious civilization would indeed pose, it is better to work with the upper-limit of the estimated probability of invasion as well as number of intelligent civilizations in our galaxy. The upper limit of the standard deviation gives a probability of invasion of 0.028%.

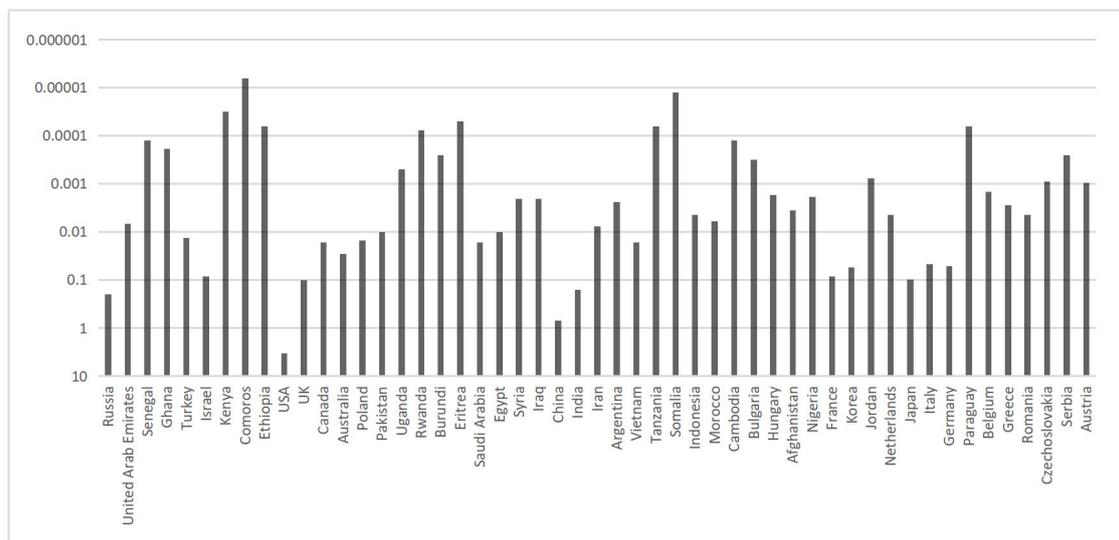

Figure 1 - Logarithmic-scale distribution of weighted probabilities of invasion by country

Now we have to search for a correlation between the global probability of invasion and the global energy consumption, which reflects technological advancement in the Kardashev Scale. In the last 50 years [see figure 2], the frequency of invasions has been decreasing at an average rate of 1.15%. Meanwhile, world energy consumption has been increasing at a steady pace of 2.24%. We can extrapolate the data to determine whether a more advanced civilization could be more or less prone to invade our planet than humanity invading theirs. With an average global consumption of 3.14E13 watts, humanity is expected to become a Type-1 civilization (1E16 watts) in 259.5 years. Such civilization has an estimated probability of 0.0014% of invading another one. Based on the upper limit of the estimated number of civilizations, which is 15,785 (Maccone, 2012), the estimated number of malicious civilizations is 0.22.

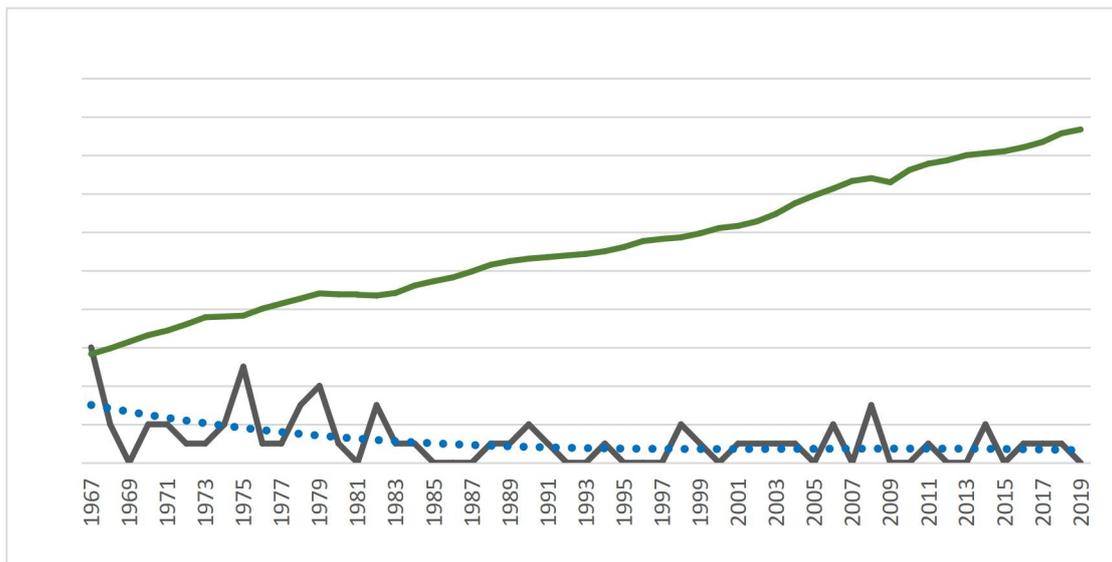

Figure 2 - Comparison between global invasion frequency (gray) and energy consumption (green) from 1967 to 2019

However, we want to calculate the probability that the exoplanet we specifically send a message to is inhabited by a malicious civilization. Not every habitable exoplanet is expected to have complex life. Based on the estimated number of potentially habitable planets in our galaxy, which is a minimum of 40 billion planets (PHL, 2022), and the upper limit of the estimated number of civilizations (15,785), the probability that the planet we message has a civilization with malicious intentions goes down to 5.52E-8 %.

We can now compare the probability of invasion of a civilization more advanced than humanity with the probability of a planet-killer asteroid collision. The impact probability of a Chicxulub-like asteroid (which led to the extinction of 75% of life) is only one every one hundred million years, or 1E-06 % (CNEOS, 2022). The probability of extraterrestrial invasion by a civilization whose planet we message is, therefore, around two orders of magnitude lower than the probability of a planet-killer e asteroid collision. Some attempts are currently being made through NASA's DART mission to deflect potentially hazardous asteroids such as Didymos B, a 160-meter asteroid, much smaller than the Chicxulub asteroid of 10 kilometers in diameter. The possibility of an asteroid colliding with Earth will always exist, including those of interstellar origin that we are not aware of.

For the probability of invasion by a messaged extraterrestrial civilization to reach the impact probability of a Chicxulub-like asteroid, we could send a maximum of 18 interstellar messages to different potentially habitable exoplanets. That limit increases when comparing it with the impact probability of smaller yet global-scale asteroids. For example, the impact probability of a 1-km asteroid that would produce a global catastrophe is 0.001% - one every 100,000 years. Such asteroid would produce an estimated number of near 2 billion deaths (CNEOS, 2022). For the probability of invasion by a messaged extraterrestrial civilization to reach such probability, we could send a maximum of 18,000 messages to different habitable planets, which would increase our chances of extraterrestrial contact while minimizing the risks.

Conclusions

An extrapolation from all the invasions that occurred on Earth in the last century shows an estimated probability of 5.52E-8 % that a messaged extraterrestrial civilization - or a faction of it - would have hostile intentions towards humanity, as well as a total of 0.22 malicious civilizations. Such a low probability is due to the fact that the estimated number of potentially habitable planets is much higher than the number of Type-1 civilizations capable of nearby interstellar travel. Moreover, there seems to be an indirect correlation between the human probability of invasion and the world's development based on energy consumption. The more developed humanity is, the less likely invasions are.

The fact that the estimated probability of extraterrestrial invasion is two orders of magnitude lower than that of a planet-killer asteroid collision should open the door to the next step, which is having an international debate to determine the conditions under which the first serious interstellar radio or laser message will be sent to a nearby potentially habitable exoplanet. It is necessary to mention that the probabilities are cumulative, which means that sending radio messages to several potentially habitable planets raises the total to the sum of all of them, which is in any case an extremely low probability. We could send up to 18,000 interstellar messages to different exoplanets and the probability of invasion by a malicious civilization would be the same as that of an Earth collision with a global-catastrophe asteroid.

# References

[1] Claudio Maccone (2012) 'The statistical Drake equation'. Link: https://link.springer.com/chapter/10.1007/978-3-642-27437-4_1

[2] CNEOS (2022) 'Earth Impact Monitoring'. Link: https://cneos.jpl.nasa.gov/sentry/

[3] SIPRI (2022) Trends in World Military Expenditure, 2021'. Link: https://www.sipri.org/sites/default/files/2022-04/fs_2204_milex_2021_0.pdf

[4] Our world in data (2022) 'How much energy does the world consume?'. Link: https://ourworldindata.org/energy-production-consumption

[5] PHL (2022) 'The Habitable Universe'. Link: https://phl.upr.edu/projects/the-habitable-universe

Annex 1

| Country | Invasions | % | Expenditure | Weight | Probability |
|---|---|---|---|---|---|
| Russia | 10 | 6,329113924 | 3,1 | 0,031 | 0,196202532 |
| United Arab Emirates | 1 | 0,632911392 | 1,07 | 0,0107 | 0,006772152 |
| Senegal | 2 | 1,265822785 | 0,01 | 0,0001 | 0,000126582 |
| Ghana | 1 | 0,632911392 | 0,03 | 0,0003 | 0,000189873 |
| Turkey | 3 | 1,898734177 | 0,7 | 0,007 | 0,013291139 |
| Israel | 11 | 6,962025316 | 1,2 | 0,012 | 0,083544304 |
| Kenya | 1 | 0,632911392 | 0,005 | 0,00005 | 3,16456E-05 |
| Comoros | 1 | 0,632911392 | 0,001 | 0,00001 | 6,32911E-06 |
| Ethiopia | 1 | 0,632911392 | 0,01 | 0,0001 | 6,32911E-05 |
| USA | 14 | 8,860759494 | 38 | 0,38 | 3,367088608 |
| UK | 5 | 3,164556962 | 3,2 | 0,032 | 0,101265823 |
| Canada | 2 | 1,265822785 | 1,3 | 0,013 | 0,016455696 |
| Australia | 3 | 1,898734177 | 1,5 | 0,015 | 0,028481013 |
| Poland | 4 | 2,53164557 | 0,6 | 0,006 | 0,015189873 |
| Pakistan | 3 | 1,898734177 | 0,53 | 0,0053 | 0,010063291 |
| Uganda | 2 | 1,265822785 | 0,04 | 0,0004 | 0,000506329 |
| Rwanda | 2 | 1,265822785 | 0,006 | 0,00006 | 7,59494E-05 |
| Burundi | 1 | 0,632911392 | 0,04 | 0,0004 | 0,000253165 |
| Eritrea | 1 | 0,632911392 | 0,008 | 0,00008 | 5,06329E-05 |
| Saudi Arabia | 1 | 0,632911392 | 2,6 | 0,026 | 0,016455696 |
| Egypt | 3 | 1,898734177 | 0,53 | 0,0053 | 0,010063291 |
| Syria | 4 | 2,53164557 | 0,08 | 0,0008 | 0,002025316 |
| Iraq | 4 | 2,53164557 | 0,08 | 0,0008 | 0,002025316 |
| China | 8 | 5,063291139 | 14 | 0,14 | 0,708860759 |
| India | 7 | 4,430379747 | 3,6 | 0,036 | 0,159493671 |
| Iran | 1 | 0,632911392 | 1,2 | 0,012 | 0,007594937 |
| Argentina | 2 | 1,265822785 | 0,19 | 0,0019 | 0,002405063 |
| Vietnam | 10 | 6,329113924 | 0,26 | 0,0026 | 0,016455696 |
| Tanzania | 1 | 0,632911392 | 0,01 | 0,0001 | 6,32911E-05 |
| Somalia | 1 | 0,632911392 | 0,002 | 0,00002 | 1,26582E-05 |
| Indonesia | 2 | 1,265822785 | 0,35 | 0,0035 | 0,00443038 |
| Morocco | 2 | 1,265822785 | 0,47 | 0,0047 | 0,005949367 |
| Cambodia | 1 | 0,632911392 | 0,02 | 0,0002 | 0,000126582 |
| Bulgaria | 1 | 0,632911392 | 0,05 | 0,0005 | 0,000316456 |
| Hungary | 3 | 1,898734177 | 0,09 | 0,0009 | 0,001708861 |
| Afghanistan | 1 | 0,632911392 | 0,56 | 0,0056 | 0,003544304 |
| Nigeria | 3 | 1,898734177 | 0,1 | 0,001 | 0,001898734 |
| France | 5 | 3,164556962 | 2,7 | 0,027 | 0,085443038 |
| Korea | 4 | 2,53164557 | 2,15 | 0,0215 | 0,05443038 |
| Jordan | 1 | 0,632911392 | 0,12 | 0,0012 | 0,000759494 |
| Netherlands | 1 | 0,632911392 | 0,7 | 0,007 | 0,00443038 |
| Japan | 6 | 3,797468354 | 2,6 | 0,026 | 0,098734177 |
| Italy | 5 | 3,164556962 | 1,5 | 0,015 | 0,047468354 |
| Germany | 3 | 1,898734177 | 2,7 | 0,027 | 0,051265823 |
| Paraguay | 1 | 0,632911392 | 0,01 | 0,0001 | 6,32911E-05 |
| Belgium | 1 | 0,632911392 | 0,23 | 0,0023 | 0,001455696 |
| Greece | 2 | 1,265822785 | 0,22 | 0,0022 | 0,00278481 |
| Romania | 3 | 1,898734177 | 0,23 | 0,0023 | 0,004367089 |
| Czechoslovakia | 1 | 0,632911392 | 0,14 | 0,0014 | 0,000886076 |
| Serbia | 1 | 0,632911392 | 0,04 | 0,0004 | 0,000253165 |
| Austria | 1 | 0,632911392 | 0,15 | 0,0015 | 0,000949367 |